\documentclass[12pt]{iopart}
\usepackage{amssymb}
\usepackage{graphicx,bm, cite,color}
\usepackage{setstack}
\usepackage{lscape}
\begin{document}
	
	\title[]{Quantum solvability of a nonlinear $\delta$-type mass profile system: Coupling constant quantization}
	\author{V. Chithiika Ruby, V. K. Chandrasekar* and M. Lakshmanan}
\address{Department of  Nonlinear Dynamics, School of Physics,
		Bharathidasan University, Tiruchirapalli - 620 024, Tamilnadu, India.}

\address{*Department of Physics, Centre for Nonlinear Science and Engineering, School of Electrical and Electronics Engineering, SASTRA Deemed University, Thanjavur-613 401, Tamilnadu, India	
}	

\begin{abstract}
In this paper, we discuss the quantum dynamics of a nonlinear system that admits temporally localized solutions at the classical level. We consider a general ordered position-dependent mass Hamiltonian in which the ordering parameters of the mass term are treated as  arbitrary. The mass function here is singular at the origin. We observe that the quantum system admits bounded solutions but importantly the coupling parameter of the system gets quantized which has also been confirmed by the semiclassical study as well.   
\end{abstract}

\maketitle

\section{Introduction}
Several studies on physical systems with position-dependent effective mass have emerged  in recent years  due to their wide applications in the study of electronic properties
of semiconductors \cite{apps}, inhomogeneous crystals, quantum dots, quantum liquids \cite{qdots,qwell,bou} and so on. The time-independent Schr\"{o}dinger equation gets generalized when the effective mass depends on the position and it is solved using both  numerical and analytical techniques. Though difficult, it is of general interest to get exact solutions for such position-dependent mass Schr\"{o}dinger equation (PDMSE) for specific potentials. Certain nonlinear systems, specifically quadratic Li\'enard type nonlinear oscillators, are found  to possess position-dependent mass  Hamiltonians. For example, Mathews-Lakshmanan oscillator and Higgs oscillator are considered to describe the dynamics of harmonic oscillators in curved space \cite{mathews-lakshmanan, higgs}. Different studies have been carried out on these systems in the literature since their introduction in the literature\cite{carinena, ball, quesne, higgs-studies}. While quantizing these position-dependent mass (PDM) quantum systems, one should consider (i) the possible choices of ordering between momentum and mass operators in their kinetic energy term and (ii) appropriate modification on the boundary conditions. The ordering may lead to Hermitian or non-Hermitian Hamiltonians. The most general ordering form had been introduced by Trabelsi et al \cite{Trabelsi}. In a recent study, it has been shown that the Mathews-Lakshmanan oscillator is exactly solvable for the general ordered form \cite{karthiga}. Motivated by the problem of ordering ambiguity of position-dependent mass Hamiltonian, two of the present authors studied the quantum dynamics of the Higgs oscillator and a $k$-dependent nonpolynomial oscillator by considering the general ordered form introduced by Trabelsi et al, in Ref. \cite{chithiika-jpa}. 

Classically both the systems, Mathews-Lakshmanan oscillator and Higgs oscillator admit non-isochronous solutions. It is recently reported that certain quadratic Li\'{e}nard type nonlinear oscillators can possess isochronous solutions as well \cite{Ajey2013}. We solved these nonlinear oscillators quantum mechanically and discussed their exact and quasi-exact solvable nature \cite{chithiika-jpco}.  It is also worth mentioning that one can also derive a conservative description for the nonlinear oscillators of position dependent linearly damped Li\'enard type systems classically. Such studies have been carried out on generalized modified Emden equation in Ref. \cite{Chandrasekar, Gladwin}. The associated Hamiltonians obtained are  non-standard. The Hamiltonian description for such a nonlinear oscillator, governed by a modified Emden equation with certain constraints on its parameters, paves a way to solve the system quantum mechanically. It is also shown that the Hamiltonian is invariant under combined coordinate reflection and time reversal transformation and exhibits linear energy spectrum  as that of the standard harmonic oscillator \cite{pt-chithiika}. 

Based on all these studies, we are here interested to study the quantum dynamics of a quadratic Li\'enard type nonlinear oscillator which shows a special behavior at its classical level. In this work, we consider such a type of nonlinear system that exhibits temporally localized solutions \cite{Ajey2013}. It is observed that the associated Hamiltonian is of the form of position-dependent mass type. The mass profile has a resemblance to a $\delta$-function form. A related model that has been used for describing electron systems in $\delta$-doped semiconductors in the Thomas-Fermi field has been shown to be quantum mechanically exactly solvable \cite{Axel}. In our work, we use a general ordering procedure to write down the appropriate quantum Hamiltonian in order to solve the underlying generalized Schr\"{o}dinger equation. We also study the role of ordering parameters on obtaining  well defined eigenfunctions as the mass function is not a continuous one here. 

In this paper, we discuss the classical solvability of the system in section 2. In section 3, we implement a semiclassical quantization rule to analyze the quantum solvability of the system and find that the coupling parameter of the system gets quantized. The system is observed as a position-dependent mass one. We consider the generalized Schr\"{o}dinger equation corresponding to a non-Hermitian ordered form to analyze the quantum solvability of the  system which is discussed in section 4. Finally, we summarize our results.

\section{A $\delta$-type mass system and its classical dynamics}
Consider a Hamiltonian of the form studied by  Tiwari et al. \cite{Ajey2013}, 
\begin{equation}
H = \frac{x^4\;p^2}{4} + \lambda x^2 
\label{hamiltonian}
\end{equation}
and the corresponding Lagrangian is 
\begin{equation}
L = \frac{\dot{x}^2}{x^4} - \lambda x^2 
\label{lagrangian}
\end{equation}
It is of the position-dependent mass form, ${\displaystyle H = \frac{p^2}{2\;m(x)}+V(x)},$ where the mass profile is of the form 
\begin{equation}
m(x) = \frac{2}{x^4} \qquad \mbox{and} \quad V(x) = \lambda x^2. 
\label{mass}
\end{equation}
Here the mass is singular at $x = 0.$

\begin{figure}[!ht]
\centering
\begin{tabular}{p{0.5\textwidth} p{0.5\textwidth}}
\hspace{1cm}\includegraphics[width=0.4\textwidth, height = 0.35\textwidth]{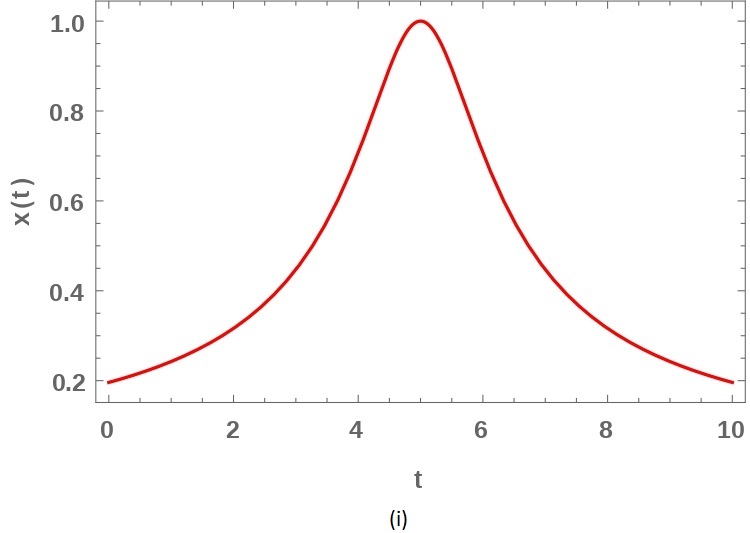} &
\includegraphics[width=0.4\textwidth, height = 0.35\textwidth]{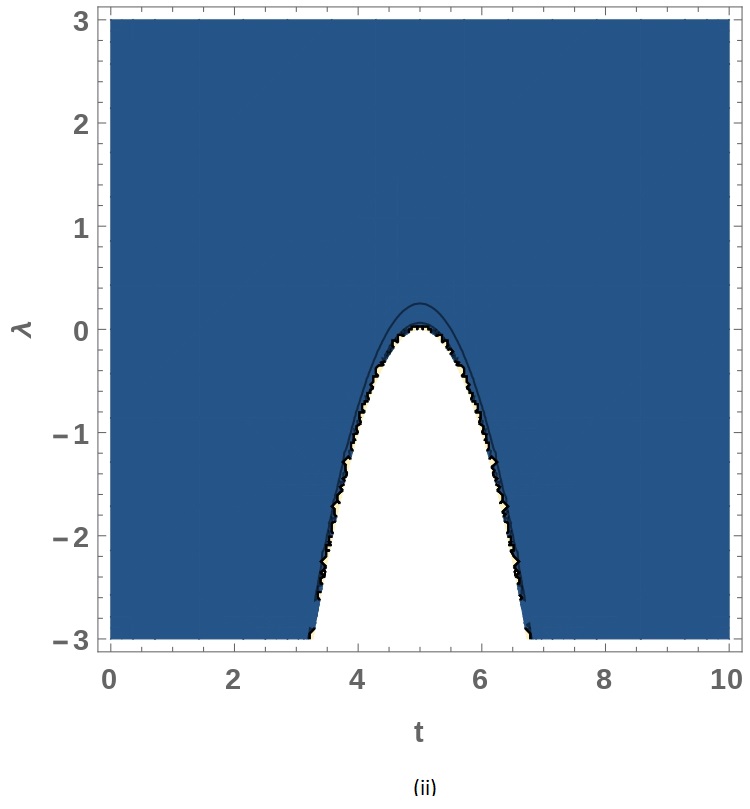}
\end{tabular}
\vspace{-0.5cm}
\caption{(i) The plot of $x(t)$ in Eq. (\ref{sol})  for $C_1 = 1,\; C_2 = -5\;$ and $\lambda = 1$ and (ii) the contour plot of $x(t)$ given in Eq. (\ref{sol}) for various values of $\lambda$ with $C_1 = 1,$ and $C_2 = -5\;$, where the blue shaded region denotes the  possible values of $\lambda$ for which the solutions are well defined and the white region denotes the values of $\lambda$ for which the solutions are singular.}
\label{classical} 
\end{figure}

The equation of motion for the Hamiltonian $H$ in (\ref{hamiltonian}) reads as 
\begin{equation}
\ddot{x} - \frac{2}{x}\;\dot{x}^2 + \lambda x^5 =0. 
\label{eqm}
\end{equation}
It can be integrated once on using the integrating factor, say ${\displaystyle \frac{2\dot{x}}{x^4}}$, as 
\begin{equation}
\frac{\dot{x}^2}{x^4} + \lambda x^2 = C_1, 
\label{xdot}
\end{equation}
where $C_1$ is an integration constant. Integrating  this equation (\ref{xdot}) once more, we find that equation (\ref{eqm}) admits the general solution,
\begin{equation}
x(t) = \frac{1}{\sqrt{\frac{\lambda}{C_1} + (C_2 + \sqrt{C_1}\;t)^2}}, 
\label{sol}
\end{equation}
where $C_2$ is the second integration constant. For $\lambda > 0$, we have a temporally localized solution. And for $\lambda < 0$, we  have  a singular solution when $t = \frac{1}{\sqrt{C_1}}\left(\sqrt{\frac{|\lambda|}{C_1}}-C_2\right)$ in which case we consider that  $C_1$ and $C_2$ are positive. The plot of $x(t)$ against $t$ is depicted in figure \ref{classical} $(i)$ for certain values of $C_1$, $C_2$ and $\lambda$. The figure \ref{classical} $(ii)$ depicts the contour plot of $x(t)$ given in Eq. (\ref{sol}) for various values of $\lambda$ with $C_1 = 1,$ and $C_2 = -5\;$.

\section{Semiclassical quantization}
\begin{figure}[hbt]
\vspace{0.5cm}
\centering
\includegraphics[width=0.4\textwidth, height = 0.4\textwidth]{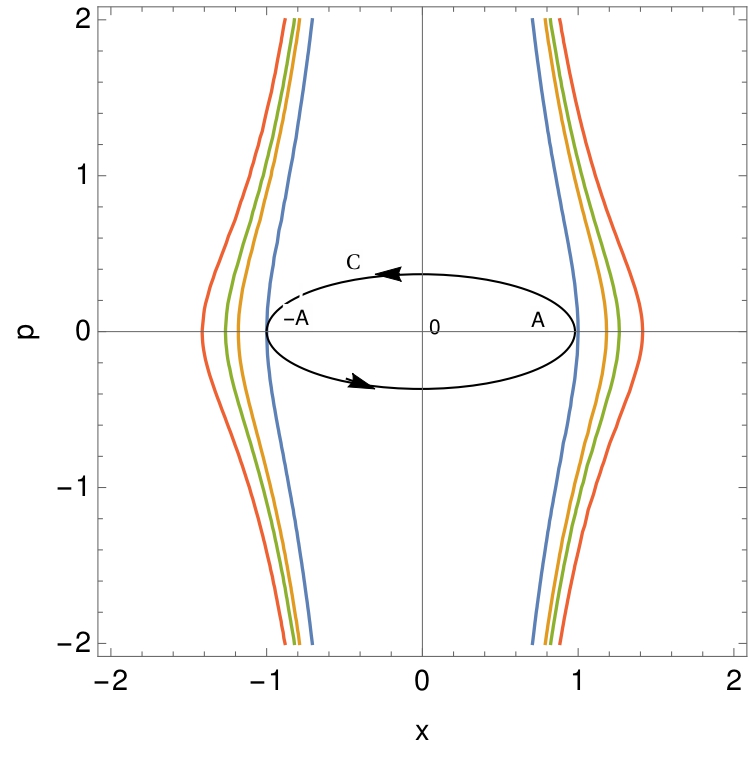}
\caption{The phase portrait of Hamiltonian (\ref{hamiltonian}) for different values of energy $E = 0.5, 0.7, 0.8,\;1$ with $\lambda = 0.5$.} \label{xp}
\end{figure}

To understand the possibility of quantization of the above  type of position-dependent mass system, we first apply the semiclassical quantization procedure to the system.  The standard leading order WKB quantization condition for the potential having two turning points is \cite{schiff}, 
\begin{equation}
\int^{x_2}_{x_1} p dx = \left(n+\frac{1}{2}\right)\hbar\;\pi, \quad n = 0, 1, 2, ..., \label{rule} 
\end{equation}
where $x_1$ and $x_2$ are the classical turning points and the conjugate momentum, $p = \sqrt{2m(x)\;(E-V(x))}$. Here, ${\displaystyle \hbar = \frac{h}{2\pi},}$ where $h$ is Planck's constant. From the Hamiltonian 
(\ref{hamiltonian}), with $H = E$, one can express the momentum as
\begin{equation}
p =   \sqrt{\frac{4 E}{x^4}-\frac{4\lambda}{x^2}}.\label{momentum} 
\end{equation}
At the turning points, say $(x_{1}, \;x_2)= (-A, A)$, the momentum is zero, which is shown in the figure \ref{xp}. Hence, from (\ref{hamiltonian}), the total energy, $H = E = \lambda A^2$ and the integral (\ref{rule}) becomes, 
\begin{equation}
2\;\sqrt{\lambda}\int^{A}_{-A} \frac{\sqrt{A^2 - x^2}}{x^2}dx = \left(n+\frac{1}{2}\right)\hbar\;\pi, \quad n = 0, 1, 2, .... \label{rule2} 
\end{equation}
To evaluate (\ref{rule2}), consider the integral 
\begin{equation}
I = \int^{A}_{-A} \frac{\sqrt{A^2 - x^2}}{x^2} dx. \label{rule3} 
\end{equation}

One can also use the classical solution $x(t)$, (vide (\ref{sol})) and evaluate the closed integral around contour $C$ (given in 
Fig.\ref{xp}) in the modified Bohr-Sommerfeld quantization rule \cite{mbohr}, 
\begin{equation}
\oint p dx = \left(n+\frac{1}{2}\right)\;h.
\label{mbr}
\end{equation}
Here, the momentum, $p(t)$, takes the form as
\begin{equation}
 p(t) = \frac{2 \dot{x}(t)}{{x(t)}^4} = -2\sqrt{C_1}(C_2 + \sqrt{C_1} t)\sqrt{\frac{\lambda}{C_1} + (C_2 + \sqrt{C_1}\;t)^2}. \label{momet}
\end{equation}

We integrate the integral (\ref{rule3}) by considering $u = \sqrt{A^2-x^2}$ and $dv = \frac{1}{x^2}dx$ and get 
\begin{eqnarray}
I &=& \left. \frac{\sqrt{A^2 - x^2}}{x} \right|^{A}_{-A}-\int^{A}_{-A}\frac{dx}{\sqrt{A^2-x^2}},\nonumber \\
 &=& 0-\left[\arcsin{\left(\frac{x}{A}\right)}\right]^{A}_{-A}, \nonumber \\
I &=& -\pi. \label{rule4} 
\end{eqnarray}

On substituting the integral (\ref{rule4}) in (\ref{rule2}), one obtains the following relation on the coupling parameter, $\lambda$, as
\begin{equation}
\lambda = \left(n+\frac{1}{2}\right)^2 \frac{\hbar^2}{4}, \qquad n = 0, 1, 2, 3, ... . \nonumber \label{lambdan}
\end{equation}

Hence, the coupling parameter $\lambda$ gets related with the quantum number $n$, as in (\ref{lambdan}).

While studying the quantum dynamics of the above type of position-dependent mass system (\ref{hamiltonian}) with a singular mass function, we meet with two difficulties: (i) how to define the configuration space and (ii) how to ensure the continuity of the eigenfunctions of the corresponding Schr\"{o}dinger equation? We proceed to incorporate these two aspects in our further study as indicated below. 

\section{Quantization: general ordered form of Hamiltonian}
We now consider the most general form of the associated Hamiltonian operator that provides a complete classification of Hermitian and non-Hermitian orderings \cite{Trabelsi}, 
\begin{eqnarray}
\hat{H} = \frac{1}{2}\sum^N_{i = 1} w_i m^{\alpha_i} \hat{p} m^{\beta_i} \hat{p} m^{\gamma_i} + V(x), 
\label{geo}
\end{eqnarray}
where  $N$ is an arbitrary positive integer and $\hat{p}$ is the one dimensional momentum operator. The ordering parameters should satisfy 
the constraints, $\alpha_i +\beta_i +\gamma_i = -1,\;i =1, 2, 3, ... N,$ and $w_i$'s  are real weights which are summed to be $1$. The above form globally connects all the Hermitian orderings and also provides a complete classification of Hermitian and non-Hermitian orderings \cite{Trabelsi}.  The operator $\hat{H}$ in (\ref{geo}) possesses $2 N$ free ordering parameters, after taking into account the above constraints.

The corresponding Hamiltonian  for the potential $V$ can be written as  
\begin{equation}
\hspace{-1.5cm} \hat{H} = \frac{1}{2}\hat{p}\frac{1}{m}\hat{p}+(\bar{\gamma} - \bar{\alpha}) \frac{i \hbar}{2} {\frac{d}{dx}} \left( \frac{1}{m}\right)\hat{p} + \frac{\hbar^2}{2}\left[\bar{\gamma} {\frac{d^2}{dx^2}}\left(\frac{1}{m}\right) + \overline{\alpha\gamma} \left(\frac{m'^2}{m^3}\right) \right] + V, \label{nhe}
\end{equation}
where ${\displaystyle \hat{p} = -i\hbar \frac{d}{dx}}$. In (\ref{nhe}), the over bar over the parameters represent their total value, $\bar{X} = \sum^{N}_{i} w_i X_i$.

The study on the effective-mass Hamiltonians for abrupt heterojunctions indicates that the single-term ordering forms of kinetic energy operator are viable candidates that ensure continuity of the associated matching conditions \cite{Morrow}. 
As the mass $m(x)$ is singular at $x=0$, we use the single term of the general ordered form of the Hamiltonian as
\begin{eqnarray}
\hat{H} = \frac{1}{2} m^{\alpha_1} \hat{p} m^{\beta_1} \hat{p} m^{\gamma_1} + V(x), \qquad \alpha_1+\beta_1+\gamma_1 = -1.  \label{single}
\end{eqnarray}
Here, we are considering non-Hermitian ordered form of the Hamiltonian (\ref{nhe}) as the non-Hermitian ordered form can be related with the Hermitian ordered form through similarity transformation \cite{chithiika-removal} as 
\begin{eqnarray}
\hat{H}_{her} &=& m^{\eta} \hat{H}\;m^{-\eta},\qquad 2\;\eta = \gamma_1 - \alpha_1, \label{geo-her1}. 
\end{eqnarray}
Consequently, for (\ref{geo-her1}) we have 
\begin{eqnarray}
\hat{H}_{her} &=& \frac{1}{2} m^{\frac{\gamma_1 +\alpha_1}{2}} \hat{p} m^{\beta_1} \hat{p} m^{\frac{\gamma_1 +\alpha_1}{2}}+ V(x).  \label{geo-her2}
\end{eqnarray}

As the non-Hermitian ordered form (\ref{nhe}) is being related with the Hermitian ordered form through similarity transformation (\ref{geo-her1}), we use the non-Hermitian ordered form of the Hamiltonian in this present work and analyze the possibility of obtaining a  complete set of solutions of the operator (\ref{nhe}). 
 
The time-independent Schr\"{o}dinger equation for the non-Hermitian ordered Hamiltonian (\ref{single}), $\hat{H}\psi = E \psi,$ can be written as   
\begin{eqnarray}
\fl\quad \psi{''} +\left({\gamma}_1-{\alpha}_1-1\right) \frac{m{'}}{m} \psi{'} + 
\left({\gamma}_1\;\frac{m{''}}{m}-\left({\alpha_1\gamma_1}+2{\gamma_1}\right)  \frac{m{'^2}}{m^2}\right) \psi + \frac{2 m}{\hbar^2}\left(E -V(x)\right)\psi  = 0, \nonumber \\
\label{SE1}
\end{eqnarray}
where ${\displaystyle ' = \frac{d}{dx}}$. 

As the above Hamiltonian depicts the dynamics of the one dimensional potential (\ref{hamiltonian}), we use the generalized position-dependent mass Schr\"{o}dinger equation  resulting from the non-Hermitian ordering (\ref{single}), to study the solvability of the system (\ref{hamiltonian}). It results that 
\begin{eqnarray}
\hspace{-0.5cm} \psi{''} +\frac{4\left(1+{\alpha}_1-{\gamma}_1\right)}{x}\;\psi{'} +\left[ \frac{4\;E}{\hbar^2\;x^4} -\frac{16{\alpha_1\gamma_1}+12{\gamma}_1+\frac{4\lambda}{\hbar^2}}{x^2}\right]\psi  = 0.  
\label{SE2}
\end{eqnarray}

By using the transformation, $\psi(x) = x^d\;\phi(x)$, where $d$ is a parameter to be determined, we can reduce the equation (\ref{SE2}) to the form 
\begin{eqnarray}
\hspace{-2.5cm} \phi{''} +\frac{2d+4\left(1+{\alpha}_1-{\gamma}_1\right)}{x}\;\phi{'} +\left[ \frac{d(d+3+4({\alpha}_1-{\gamma}_1))-\left(16{\alpha_1\gamma_1}+12{\gamma}_1+\frac{4\lambda}{\hbar^2}\right)}{x^2}+\frac{4\;E}{\hbar^2\;x^4}\right]\phi  = 0. \nonumber \\
\label{SE3}
\end{eqnarray}

We further use the transformation, ${\displaystyle g(x) = \frac{-1}{2\;x}}$, so that Eq. (\ref{SE3}) can be rewritten as 
\begin{eqnarray}
\hspace{-1cm} g^2\phi_{gg} +2g\;\left[\left(2{\gamma}_1-2{\alpha}_1-1-d\right)\right]\;\phi_g +\left[ d(d+3+4({\alpha}_1-{\gamma}_1)) -\left(16{\alpha_1\gamma_1}+12{\gamma_1}+\frac{4\lambda}{\hbar^2}\right) \right. \nonumber \\
\hspace{5cm}\left.+\frac{16\;E}{\hbar^2}\;g^2\right]\phi  = 0, \hspace{-2cm}
\label{SE4}
\end{eqnarray}
where ${\displaystyle \phi_g = \frac{d\;\phi}{dg}}$. 

In order to map Eq. (\ref{SE4}) to the known form, we again use the transformation, 
\begin{eqnarray}
\tau &=&\frac{4\sqrt{E}}{\hbar}\;g, \\
\hspace{-2.5cm}\mbox{with}\nonumber \\
d &=& 2{\gamma}_1-2{\alpha}_1 - \frac{3}{2}, \label{dvalue} 
\end{eqnarray}
to transform equation (\ref{SE4}) as 
\begin{eqnarray}
{\tau}^2\phi_{\tau \tau} + \tau\;\phi_{\tau} +\left( {\tau}^2 - \nu^2 \right)\phi  = 0,
\label{SE5}
\end{eqnarray}
where 
\begin{eqnarray}
\nu^2 = \left(2{\alpha}_1+2{\gamma}_1+\frac{3}{2}\right)^2 +\frac{4\lambda}{\hbar^2}. \label{nu}
\end{eqnarray}

Eq. (\ref{SE4}) is of the form of Bessel's differential equation. Hence, the corresponding general solution is 
\begin{equation}
\phi_{\nu} (\tau)= C J_{\nu}(\tau) + D Y_{\nu}(\tau), \label{sol4}
\end{equation}
where $J_{\nu}(\tau)$ and  $Y_{\nu}(\tau)$ are the first and second kind of Bessel polynomials \cite{ryzik} and $C$ and $D$ are arbitrary constants. Now we can obtain the general solution for the equation (\ref{SE2}) for the region $x \in (0, \infty)$ as 
\begin{equation}
\psi_{\nu}(x) = \psi^{(+)}_{\nu}(x) = x^{d}\left[ C J_{\nu}\left(\frac{2\sqrt{E}}{\hbar\; x}\right) +D Y_{\nu}\left(\frac{2\sqrt{E}}{\hbar\; x}\right)\right], \quad x \in (0, \infty). \label{sol2}
\end{equation}

And we can write down the  general solution for the region $x < 0$, as 
\begin{equation}
\psi^{(-)}_{\nu}(x) = (-|x|)^{d}\left[ \tilde{C} J_{\nu}\left(\frac{2\sqrt{E}}{\hbar\; x}\right) +\tilde{D} Y_{\nu}\left(\frac{2\sqrt{E}}{\hbar\; x}\right)\right], \quad x \in (-\infty, 0).  \label{sol2-}
\end{equation}
where $\tilde{C}$ and $\tilde{D}$ are arbitrary constants and $d$ (vide Eq. (\ref{dvalue})). 

Here we are interested to derive bounded solutions for the system (\ref{hamiltonian}) and so analyze the boundary conditions for the Bessel  polynomials.

By choosing $d = 2\gamma_1-2\alpha_1-1$, equation (\ref{SE4}) can now be reduced to the constant mass Schr\"{o}dinger equation as 
\begin{eqnarray}
\hspace{-1cm} \phi_{gg} +\left[\frac{16\;E}{\hbar^2}-\frac{\frac{4\lambda}{\hbar^2}+\left(2\alpha_1+2\gamma_1+2\right)\left(2\alpha_1+2\gamma_1+1)\right)}{g^2}\right]\phi  = 0.  
\label{SE-pct}
\end{eqnarray}
This equation can also be deduced by means of a point canonical transformation method, which relates the PDM Schr\"{o}dinger equation with the canonical form of constant mass Schr\"{o}dinger equation and it is a widely used method in solving position-dependent mass Schr\"{o}dinger equations \cite{pct}. The potential of (\ref{SE-pct}), $U(g) \propto \frac{1}{g^2}$, is similar to the effective potential that  arose while studying the Efimov effect in the quantum three body system that describes the dynamics of two heavy particles interacting through a light particle \cite{Efimov}.

\subsection{Boundary conditions}
In Eq. (\ref{sol2}), when $x\rightarrow \infty$ the polynomials $J_{\nu}$ become zero for positive values of $\nu$ and become complex infinity for $\nu < 0$. And $Y_{\nu}$ becomes $\infty$ provided $\nu \neq 0.$ Hence, we take $D = 0$ and $\nu > 0$ to get the solutions which are bounded as $x \rightarrow \infty.$ 

To proceed further, we now expand (\ref{sol2}) around $x = \infty$, 
 \begin{equation}
\psi^{(+)}_{\nu}(x) = C\; x^{d} \; J_{\nu}\left(\frac{2\sqrt{E}}{\hbar\; x}\right)  \approx_{x \to \infty}  \frac{C}{\Gamma{(\nu+1)}}\left(\frac{E}{\hbar^2}\right)^{\nu/2}\;x^{d-\nu}. \label{sol3}
\end{equation}
The boundary condition on $\psi^{(+)}_{\nu}(x)$ at $x \rightarrow \infty$ fixes a constraint $d - \nu < 0$. As $\nu > 0$, the value of $d$ fixes the lower bound of $\nu.$

Secondly we analyze the bounded nature of $\psi^{(+)}_{\nu}(x)$ at $x = 0.$ When $x$ approaches  zero,  ${\displaystyle J_{\nu}\left(\frac{2\sqrt{E}}{\hbar\; x}\right)}$ oscillates vastly as  ${\displaystyle \frac{2\sqrt{E}}{\hbar\; x}}$ goes to $\infty.$ On expanding near zero, we obtain 
\begin{equation}
\fl \quad \psi^{(+)}_{\nu}(x) =C\; x^{d} \; J_{\nu}\left(\frac{2\sqrt{E}}{\hbar\; x}\right)  \approx_{x-> 0} C\;\sqrt{\frac{\hbar\;x}{\pi\;\sqrt{E}}}x^d\;\cos\left(\frac{2\sqrt{E}}{\hbar\;x}-\frac{\pi}{2}\left(\nu+\frac{1}{2}\right)\right). \label{sol4}
\end{equation}

Here we use the squeeze theorem which states that if a function $g(x)$ is squeezed between the functions $f(x)$ and $h(x)$ near a point $a$ and if $f(x)$ and $h(x)$ have the same limit $L$ at the point $a$, then $g(x)$ is trapped and will be forced to have also the same limit $L$ at $a$ \cite{squeeze}. 
Since near $x=0$, the cosine function is not well defined as $-1 \le \cos\left(\frac{2\sqrt{E}}{\hbar\;x}-\frac{\pi}{2}\left(\nu+\frac{1}{2}\right)\right) \le 1$, in accordance with the squeeze theorem, if we consider the functions, $f(x) = \sqrt{x}$ and $h(x) = -\sqrt{x}$, then  the $\lim_{x-> 0}\sqrt{x}=0$ makes $\lim_{x->0}\;\sqrt{x}\cos\left(\frac{2\sqrt{E}}{\hbar\;x}-\frac{\pi}{2}\left(\nu+\frac{1}{2}\right)\right) = 0$. 
\begin{itemize}
\item Hence, for the values of $d < 0$, the solutions $\psi^{(+)}_{\nu}(x)$ are not well defined near zero. It restricts that $d \ge 0.$ 
\item But we have  $d - \nu < 0$ which fixes the lower bound of $\nu$. To consider the lower bound value of $\nu$ as the least of the value of $\nu$, we consider $d =0$. 
\end{itemize}

Hence, the eigenfunction, Eq. (\ref{sol2}) becomes
\begin{eqnarray}
\psi^{(+)}_{\nu}(x) = C J_{\nu}\left(\frac{2\sqrt{E}}{\hbar\; x}\right), \quad x \in (0, \infty). \label{sol2l}
\end{eqnarray}
Similarly, the eigenfunction, Eq. (\ref{sol2-}) takes the form, 
\begin{eqnarray}
\psi^{(-)}_{\nu}(x) = \tilde{C} J_{\nu}\left(\frac{2\sqrt{E}}{\hbar\; x}\right), \quad x \in (-\infty, 0).  \label{sol2r}
\end{eqnarray}

We also consider that $\nu > 0$ from the fact that the Bessel functions $J_{\nu}(0)$ are not well defined at $\nu = 0$.

\subsection{Parity} 
Now we use the parity condition on $J_{\nu}$. The solution (\ref{sol2r}), defined in  the region $x \in (-\infty, 0)$, may be symmetric or anti-symmetric with  $\psi^{(+)}_{\nu}(x)$.  Consider a point $\epsilon$ near $x = 0$, then we have  
\begin{eqnarray}
\hspace{0.5cm} \left. \tilde{C} \psi^{(-)}_{\nu}(x) \right|_{x = -\epsilon} &=& \left. C \psi^{(+)}_{\nu}(x)\right|_{x=\epsilon}, \label{match1} 
\end{eqnarray}
and so 
\begin{eqnarray}
\fl \qquad \qquad \left(C - (-1)^{\nu}\;\tilde{C}\right)\;J_{\nu}\left(\frac{2\sqrt{E}}{\hbar\; \epsilon}\right) & = & 0. 
\end{eqnarray}
The odd parity determines $\nu =1, 3, 5, ... , $ odd integers, and so $\tilde{C} = -C,$ whereas even parity leads to $\nu = 2, 4, ... ,$ even integers, so that  $\tilde{C} = C$. 
  
Hence, the parity condition fixes
\begin{equation}
\nu = n, \quad n = 1, 2, 3, ... . 
\label{nu-n}
\end{equation} 

As a result, we find that the coupling parameter (\ref{nu}) is now related with the quantum number `$n$' as 
\begin{eqnarray}
\lambda = \left(n^2 - \left(2{\alpha}_1+2{\gamma}_1+\frac{3}{2}\right)^2\right)\frac{\hbar^2}{4}, \quad n = 1, 2, 3, ...  \label{lambda-n}
\end{eqnarray}
 and so it is quantized which has also been confirmed by the semiclassical quantization method, vide  Eq. (\ref{lambdan}). 

Hence, the bound states from (\ref{sol2l}) and (\ref{sol2r}) become
\begin{eqnarray}
\hspace{0.1cm} \psi^{(+)}_{n}(x) &=&  C J_{n}\left(\frac{2\sqrt{E}}{\hbar\; x}\right),  \quad x \in (0, \infty) \quad n=1,2, 3, ....\label{sol_p}\\
\hspace{0.1cm} \psi^{(-)}_{n}(x) &=&  C (-1)^n J_{n}\left(\frac{2\sqrt{E}}{\hbar\; |x|}\right),  \quad x \in (-\infty, 0) \quad n=1,2, 3, ....\label{sol_m}
\end{eqnarray}

The parity nature of the eigenfunctions (\ref{sol_p}) and (\ref{sol_m}) restricts the coupling parameter to take discrete values, that is expressed in terms of quantum number $'n'$ in (\ref{lambda-n}). Subsequently we analyze the energy eigenvalues in the following subsection.

\subsubsection{\bf Energy:}
As $m(x) = \frac{2}{x^4}$ is singular at $x=0$, the eigenfunctions $\psi^{(\pm)}_n(x)$ (vide Eq. (\ref{sol_p}) and Eq.(\ref{sol_m}))  are  restricted to be zero at that point $x=0$, that is 
\begin{eqnarray}
\lim_{x-> 0}\psi^{(\pm)}_n(x) = 0. 
\end{eqnarray}
Consequently, we have 
\begin{eqnarray}
\lim_{x-> 0}\sqrt{\frac{\hbar x}{\pi \sqrt{E}}}\;\cos\left(\frac{2\sqrt{E}}{\hbar\;x}-\frac{\pi}{2}\left(n+\frac{1}{2}\right)\right)= 0. 
\end{eqnarray}
The above relation establishes that the energy eigenvalues are continuous, while the coupling parameter $\lambda$ is quantized as in Eq. (\ref{lambda-n}). 

\subsection{Normalizability condition of the states (\ref{sol_p}) and (\ref{sol_m}):}
As the non-Hermitian ordered form of the Hamiltonian can be related with the Hermitian ordered form through similarity transformation, one can express the normalization condition for non-Hermitian ordered Hamiltonian as \cite{chithiika-removal}, 
\begin{eqnarray}
1 = \langle \psi^{(\pm)}_{n} m^{2\eta}|\psi^{(\pm)}_n\rangle, \label{nn}
\end{eqnarray}
where $\eta = \frac{\gamma_1-\alpha_1}{2}$. On substituting (\ref{sol_p}) in (\ref{nn}), we can get 
\begin{eqnarray}
 1 = C^2 2^{\gamma_1 - \alpha_1}\;\int^{\infty}_0 \frac{1}{x^{4\gamma_1 - 4\alpha_1}}J_n\left(\frac{2\sqrt{E}}{\hbar\;x}\right)J_n\left(\frac{2\sqrt{E'}}{\hbar\;x}\right)dx. \label{integral}
\end{eqnarray}
As $d =0$, we have $\gamma_1-\alpha_1 = \frac{3}{4}$. By applying a simple transformation $\rho = \frac{1}{x}$ to (\ref{integral}), we can get
\begin{eqnarray}
1 &=& C^2 2^{3/4}\;\int^{\infty}_0 \rho \; J_n\left(\frac{2\sqrt{E}\;\rho}{\hbar}\right)J_n\left(\frac{2\sqrt{E'}\;\rho}{\hbar}\right)d\rho. \label{integral1}
\end{eqnarray}
 On using the identity, 
 \begin{eqnarray}
 \int^{\infty}_{0} k J_n(k a) J_n(k b) dk = \frac{1}{a}\delta(b-a), \qquad n = 0, 1, 2, ... , 
 \end{eqnarray}
 we can obtain the condition
\begin{equation}
1 = \frac{C^2 2^{3/4}\hbar}{2\sqrt{E}} \delta\left(\frac{2\sqrt{E'}}{\hbar}-\frac{2\sqrt{E}}{\hbar}\right)
\end{equation}
where $\delta(a-b)$ is the Dirac delta function which becomes infinity when $a = b$, otherwise it has zero value. 

We now obtain, 
\begin{equation}
C = \left(\frac{\sqrt{2\;E}}{\hbar \; \delta\left(\frac{2\sqrt{E}}{\hbar}-\frac{2\sqrt{E'}}{\hbar}\right)}\right)^{1/2}. \label{norm}
\end{equation}
As the energy eigenvalue of the system is arbitrary and continuous, we have obtained the normalization constant in terms of Dirac delta function. This is analogous to the quantization of a free particle on a cone studied recently by Kowalski et al. \cite{kowalski}.  

Hence, we obtained the bounded states (\ref{sol2}) in both the regions, $x\in (0, \infty)$ and $x \in (-\infty, 0)$, as  
\begin{eqnarray}
\hspace{0.1cm} \psi^{(\pm)}_{n}(x) = C  J_{n}\left(\frac{2\sqrt{E}}{\hbar\; x}\right), \quad n=1,2, 3, ....  \label{solf}
\end{eqnarray}

\begin{figure}[h]
\vspace{0.5cm}
\begin{tabular}{p{0.5\textwidth} p{0.5\textwidth}}
\includegraphics[width=0.4\textwidth, height = 0.4\textwidth]{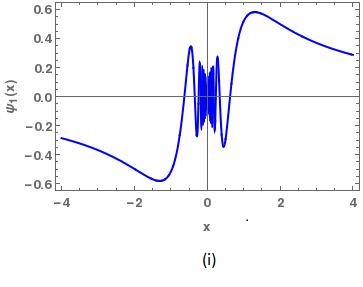} &
\includegraphics[width=0.4\textwidth, height = 0.4\textwidth]{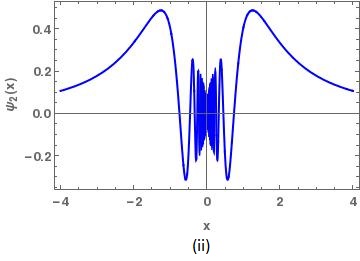}
\end{tabular}
\caption{The plots of (i)  $\psi_1(x)$ and (ii) $\psi_2(x)$ in Eq. (\ref{solf}).} 
\label{eigen}
\end{figure}

The first two states (unnormalized) are plotted in the figure \ref{eigen}. 

One can reinterpret the normalization condition, 
\begin{equation}
1 = \int^{\infty}_{-\infty}\psi^*_n(x)\psi_n(x)\;dx,
\label{norm-re}
\end{equation}
 by omitting the singular region $(-\epsilon, \epsilon)$ and reconsidering the integral (\ref{nn}) by 
\begin{eqnarray}
1 = 2 C^2 \int^{\infty}_{\epsilon} \frac{1}{x^3}\; J_{n}\left(\frac{2\sqrt{E}}{\hbar\; x}\right)J_{n}\left(\frac{2\sqrt{E'}}{\hbar\; x}\right)\;dx,  \label{inte1}
\end{eqnarray}
in which we considered (\ref{solf}). 

Let ${\displaystyle \frac{1}{x} = \rho}$. The integral (\ref{inte1}) becomes 
\begin{eqnarray}
1 = 2 C^2 \int^{1/\epsilon}_{0} \rho\;J_{n}\left(\frac{2\sqrt{E}}{\hbar\;} \rho\right)J_{n}\left(\frac{2\sqrt{E'}}{\hbar\; }\rho\right)d\rho. \label{inte2}
\end{eqnarray}
Now we use the identity \cite{arfen}
\begin{equation}
\int^a_0 \rho J_{\nu}\left(\alpha_{\nu m}\frac{\rho}{a}\right)J_{\nu}\left(\alpha_{\nu m}\frac{\rho}{a}\right)\;d\rho = \frac{a^2}{2}\;\left[J_{\nu+1}(\alpha_{\nu m})\right]^2\delta_{nm}, \label{iden2}
\end{equation}
where $\delta_{nm}$ is Kronecker delta function that takes the value $1$ when $n=m$ otherwise it takes zero. Here, $\alpha_{\nu m}, \;m = 1, 2, 3, ... \infty,$ is the $m^{th}$ zero of the Bessel function $J_{\nu}$, that is $J_{\nu}(\alpha_{\nu m}) = 0.$

The integral (\ref{inte2}) now becomes 
\begin{eqnarray}
1 = \frac{C^2}{\epsilon^2}\left[J_{n+1}\left(\frac{2\sqrt{E^{N}_n}}{\hbar\;\epsilon}\right)\right]^2 \label{int-3}
\end{eqnarray}
which makes the energy eigenvalues to take the values, 
\begin{equation}
E^{N}_n = \frac{\hbar^2}{4}\;j^{N^2}_n\;\epsilon^2, \qquad \epsilon \neq 0,\label{energy}
\end{equation}
where $j^N_n, \; N=1, 2, 3, ...\infty, \; n = 1, 2, 3, ...$ are zeroes of the Bessel function, $J_n.$ 
The normalization constant reads as 
\begin{eqnarray}
C^N_n= \frac{\epsilon}{J_{n+1}\left(\frac{2\sqrt{E^{N}_n}}{\hbar\;\epsilon}\right)}.\label{int-4}
\label{normalization}
\end{eqnarray}

The normalized eigenstates, vide (\ref{energy}) and (\ref{normalization}), can be written  as
\begin{equation}
\hspace{0.1cm} \psi^N_{n}(x) =  C^N_n  J_{n}\left(\frac{2\sqrt{E^N_n}}{\hbar\; x}\right), \quad n=1,2, 3, .... \quad N = 1, 2, 3, ..., \quad \epsilon \neq 0. \label{solf-norm}
\end{equation}   

We have observed that one can possibly obtain the normalized eigenfucntions with the corresponding eigenvalues by restricting the motion of the particle around a point near to the origin $\epsilon\;(\neq 0).$

\subsection{Hermitian ordering}
In the previous section, we considered non-Hermitian ordered Hamiltonian (\ref{single}) and solved the corresponding generalized Schr\"{o}dinger equation that resulted in the general solution (\ref{solf}). In this sub-section, we discuss about the solution of the Hermitian ordered form of the Hamiltonian (\ref{geo-her2}). 
\begin{eqnarray}
\hspace{0.5cm}\hat{H}_{her} = \frac{1}{2} m^{\frac{\gamma_1 +\alpha_1}{2}} \hat{p} m^{\beta_1} \hat{p} m^{\frac{\gamma_1 +\alpha_1}{2}}+ V(x). \nonumber\hspace{5.1cm}\mbox{(\ref{geo-her2})}
\end{eqnarray}

Instead of solving the Schr\"{o}dinger equation corresponding to the Hermitian ordered Hamiltonian (\ref{geo-her2}), we can obtain the solution from the relation (\ref{geo-her1}) that relates the non-Hermitian ordered form (\ref{nhe})  with the Hermitian ordered form through similarity transformation. 
\begin{eqnarray}
\hat{H}\psi &=&  m^{-\eta} \hat{H}_{her} m^{\eta}\psi, \qquad 2\eta =\gamma_1-\alpha_1. \label{new}
\end{eqnarray}
Let $m^{\eta}\psi = \phi$. As we have $2\eta =\gamma_1-\alpha_1 = \frac{3}{2}$ from $d=0$,  we can write down the solution for (\ref{geo-her2}) from (\ref{solf}), 

\begin{equation}
\hspace{0.1cm} \phi_{n}(x) =  C m^{\eta} J_{n}\left(\frac{2\sqrt{E}}{\hbar\; x}\right) =  C\;x^{-3/2} J_{n}\left(\frac{2\sqrt{E}}{\hbar\; x}\right), \quad n=1,2, 3, ...,  \label{solf-her}
\end{equation}
where the normalization constant $C$ is the same as obtained in (\ref{norm}). The solution (\ref{solf-her}) is singular at $x=0$. Hence, for the system (\ref{hamiltonian}), the non-Hermitian ordered form (\ref{single}) only yields bounded solutions (\ref{solf}). 

\section{Conclusion}
In this work, we considered a nonlinear system of the quadratic Li\'enard type which admits temporally localized solutions at the classical level. Depending upon the positive and negative values of the coupling parameter $\lambda$, the solution is well defined 
or has a singular value in its domain. To start with, we implemented the WKB quantization condition which ensures that the coupling parameter $\lambda$ would be quantized. While studying the quantum dynamics of the system, we considered a single term of the general ordered position-dependent mass Hamiltonian as the  mass function which is singular at the origin and solved the underlying Schr\"{o}dinger equation. We observed that the quantum system admits bounded solutions. Specifically, we find that the coupling parameter of the system gets quantized. We believe that such an observation is quite new to the literature as far as the quantization is concerned.  The position dependent mass with $\delta$-type mass profile considered in this paper may find application in the field of semiconductor physics, as in the case of Thomas-Fermi potential with $\delta-$doped semiconductor \cite{Axel}. We believe that our study widens the scope of quantizing other solvable classical nonlinear oscillators exhibiting novel dynamical features in a broader  sense.

\section*{Acknowledgment} 
VC wishes to acknowledge DST for the financial support of the project (No. SR/WOS-A/PM-64/2018(G)) under Women Scientist Scheme A.  ML acknowledges the financial support under a DST-SERB National Science Chair position. The work of VKC is supported by SERB-DST-MATRICS (No. MTR/2018/000676).


\section*{References}
\begin{thebibliography}{11}
\bibitem{apps}
Bastard G 1992 {\it Wave Mechanics Applied to Semiconductor Heterostructures} (Les, Editions de Physique). 

\bibitem{qdots}   
Gora T and Williams F 1969 \emph{Phys. Rev.} {\bf 177} 1179; 
Marrow R A 1985 \emph{Phys. Rev. B} {\bf 27} 2294; 1987 {\bf 36} 4836. 

\bibitem{qwell}
Serra L I and Lipparani E 1997 \emph{Europhys. Lett.} {\bf 40} 667; 
Harrison P 2000 {\it Quantum Wells, Wires and Dots} (United Kingdom, John Wiley and Sons)

\bibitem{bou}
L\'{e}vy-Leblond J -M 1992 \emph{Eur. J. Phys.} {\bf 13} 215-218

\bibitem{mathews-lakshmanan}
Mathews P M and Lakshmanan M 1974 \emph{Q. Appl. Math.} {\bf 32} 215; 
Mathews P M and Lakshmanan M  1975 \emph{Nuovo Cimento A} {\bf 26}  299
 
\bibitem{higgs}
Higgs P W 1979 \emph{J. Phys. A: Math. Gen.} {\bf 12}  309;
Leemon H I 1979 \emph{J. Phys. A: Math. Gen.} {\bf 12} 489


\bibitem{ball} 
Ballesteros A and Herranz F J 2007 \emph{J.  Phys.  A:  Math.Theor.} {\bf 40} F51-F59;
Ballesteros A and Herranz F J 2009 \emph{J. Phys. A:Math. Theor.} {\bf 42}  245203 

\bibitem{quesne}
Quesne C 2016 \emph{Euro-physics Letters} {\bf 114} 10001; 
Quesne C 2015 \emph{Phys. Lett. A} {\bf 379} 1589-93; 
Schulze-Halberg A 2015 \emph{Eur. Phys. J. Plus} {\bf 130} 1-10. 

\bibitem{carinena}
Cari\~{n}ena J F, Ra\~{n}ada M F, Santander M and Senthilvelan M 2004 \emph{Nonlinearity} {\bf 17} 1941; 
Cari{\~n}ena J F, Ra\~{n}ada M F and Santander M 2007 \emph{Annals of Physics} {\bf 322} 434; 
Cari{\~n}ena J F,  Ra{\~n}ada M F  and  Santander M 2017 \emph{J. Phys. A: Math. Theor.} {\bf 50} 465202; 
Cari\~{n}ena J F,  Ra{\~n}ada M F  and  Santander M 2012 \emph{J.Math. Phys.} {\bf 53} 102109

\bibitem{higgs-studies}
Hakobyan T, Neressian A and Yeghilkyan V 2009 \emph{J. Phys. A: Math. Theor.} {\bf 42} 205206; 
Mohammadi V, Aghaei S and Chenaghlou A 2016 \emph{Int. J. Mod. Phys. A} {\bf 31} 1650190 


\bibitem{Trabelsi}
Trabelsi A, Madouri F, Merdaci A and Almatar A 2013 {\it Classification scheme for kinetic energy operators  with position-dependent mass} 
arXiv.org:1302.3963v1


\bibitem{karthiga}
Karthiga S, Chithiika Ruby V, Senthilvelan M and Lakshmanan M 2017 \emph{J. Math. Phys.} {\bf 58} 102110

\bibitem{chithiika-jpa}
Chithiika Ruby V and Lakshmanan M 2021 \emph{J. Phys. A: Math. Theor.} {\bf 54}  385301

\bibitem{Ajey2013}
Tiwari A K, Pandey S N, Senthilvelan M and Lakshmanan M 2013 \emph{J. Math. Phys.} {\bf 54} 053506

\bibitem{chithiika-jpco}
Chithiika Ruby and Lakshmanan M 2021 \emph{J. Phys. A. Commun.} {\bf 5} 065007

\bibitem{Chandrasekar}
Chandrasekar V K, Senthilvelan M and Lakshmanan M 2005 \emph{Phys. Rev. E} {\bf 72} 066203

\bibitem{Gladwin}
Gladwin Pradeep R, Chandrasekar V K, Senthilvelan M and Lakshmanan M 2009 \emph{J. Math. Phys.} {\bf 50} 052901

\bibitem{pt-chithiika}
Chithiika Ruby V, Senthilvelan M and Lakshmanan M 2012 \emph{J. Phys. A: Math. Theor.} {\bf 45} 382002 

\bibitem{Axel}
Axel Schulze-Halberg,  Jesus Garcia-Ravelo, Christian Pacheco-Garcia, Jose Juan Pena Gil 2013 \emph{Annals of Physics} {\bf 333} 323-334

\bibitem{schiff}
Schiff L I 2010 {\it Quantum Mechanics} (TATA McGraw-Hill, New York) 

\bibitem{mbohr}
Marinov M S and  Popov V S 1975 \emph{J. Phys. A: Math. Gen.} {\bf  8} 1575


\bibitem{Morrow}
Morrow R A and  Brownstein K R 1984 \emph{Phys. Rev. B} {\bf 30} 678-680

\bibitem{chithiika-removal}
Chithiika Ruby V,  Chandrasekar V K, Senthilvelan M and Lakshmanan M 2015 \emph{J. Math. Phys.} {\bf 56} 012103


\bibitem{ryzik}
Gradshteyn  I S and Ryzhik I M 1980  {\it Table of Integrals, Series and Products} (Academic Press, New York).


\bibitem{pct}
Aktas M and Sever R 2008 \emph{J.~Math.~Chemistry} {\bf 43}  1;
Jia C, Yi  L and Sun Y 2008 \emph{J.~Math.~Chemistry} {\bf43} 435

\bibitem{Efimov}
Fonseca A C, Redish E F and Shanley P E 1979 \emph{Nuclear Physics} {\bf A320} 273-288. 

\bibitem{squeeze}
Sohrab H Houshang  2003 {\it Basic Real Analysis} (Springer, New York)

\bibitem{kowalski}
Kowalski K 2013 {\emph{Annals of Physics}} {\bf 329} 146-157

\bibitem{arfen}
 Arfken G B and Weber H J 2005 {\it Mathematical Methods for Physicists} (Elsevier Academic Press, USA)
\end{thebibliography}
\end{document}